\documentclass[conference]{IEEEtran}
\IEEEoverridecommandlockouts
\usepackage{cite}
\usepackage{amsmath,amssymb,amsfonts}
\usepackage{algorithmic}
\usepackage{graphicx}
\usepackage{textcomp}
\usepackage{xcolor}
\usepackage{lipsum}
\usepackage{multirow}
\usepackage{diagbox}
\usepackage{float}
\usepackage{url}
\usepackage[ruled]{algorithm2e}

\def\BibTeX{{\rm B\kern-.05em{\sc i\kern-.025em b}\kern-.08em
    T\kern-.1667em\lower.7ex\hbox{E}\kern-.125emX}}
\begin{document}

    \title{A Comparative Study of Offline Models and Online LLMs in Fake News Detection

}

\author{\IEEEauthorblockN{1\textsuperscript{st} Ruoyu Xu}
\IEEEauthorblockA{\textit{Department of Computer Science} \\
\textit{Texas Tech University}\\
Lubbock, TX 79409, USA\\
ruoyxu@ttu.edu}
\and
\IEEEauthorblockN{2\textsuperscript{nd} Gaoxiang Li}
\IEEEauthorblockA{\textit{Department of Computer Science} \\
\textit{Texas Tech University}\\
City, Country \\
gaoli@ttu.edu}

}

\maketitle

\begin{abstract}
Fake news detection remains a critical challenge in today’s rapidly evolving digital landscape, where misinformation can spread faster than ever before. Traditional fake news detection models often rely on static datasets and auxiliary information, such as metadata or social media interactions, which limits their adaptability to real-time scenarios. Recent advancements in Large Language Models (LLMs) have demonstrated significant potential in addressing these challenges due to their extensive pre-trained knowledge and ability to analyze textual content without relying on auxiliary data. However, many of these LLM-based approaches are still rooted in static datasets, with limited exploration into their real-time processing capabilities. This paper presents a systematic evaluation of both traditional offline models and state-of-the-art LLMs for real-time fake news detection. We demonstrate the limitations of existing offline models, including their inability to adapt to dynamic misinformation patterns. Furthermore, we show that newer LLM models with online capabilities, such as GPT-4, Claude, and Gemini, are better suited for detecting emerging fake news in real-time contexts. Our findings emphasize the importance of transitioning from offline to online LLM models for real-time fake news detection. Additionally, the public accessibility of LLMs enhances their scalability and democratizes the tools needed to combat misinformation. By leveraging real-time data, our work marks a significant step toward more adaptive, effective, and scalable fake news detection systems.

\end{abstract}

\begin{IEEEkeywords}
Fake news detection, LLMs
\end{IEEEkeywords}

\section{Introduction}


The exponential growth of the Internet and the widespread adoption of social media platforms such as Twitter and Facebook have revolutionized the dissemination of news, making it more decentralized and rapid than ever before. However, this shift has also turned these platforms into grounds for the spread of fake news, posing significant threats to public trust and societal stability. A 2023 report by the Reuters Institute for the Study of Journalism highlighted the persistent issue of misinformation, with social media playing a central role in its proliferation \cite{reuters2023digitalnews}. Despite ongoing efforts to curb the spread of false information, the sheer scale and speed at which fake news propagates remain formidable challenges. This underscores the urgent need for effective and adaptive detection methods capable of operating in real-time to counter the evolving nature of misinformation.


Traditional approaches to fake news detection have primarily relied on offline models trained on historical datasets \cite{}. While these models have proven effective in identifying patterns associated with previously encountered fake news, they suffer from a critical limitation: their static nature prevents them from adapting to the rapidly changing landscape of misinformation. This limitation has been substantiated by our evaluations, which demonstrate that as fake news evolves in both content and dissemination strategies, offline models increasingly struggle to maintain their effectiveness. This challenge is particularly pronounced in real-time scenarios, where new narratives can emerge and spread at unprecedented speeds, further diminishing the accuracy and reliability of these models.

As fake news continues to evolve, both in terms of content and dissemination strategies, traditional offline models increasingly struggle to maintain their effectiveness. This challenge is particularly pressing in real-time scenarios, where new narratives can emerge and spread at unprecedented speeds. A significant issue lies in the distributional shift between the static, historical data on which these models are trained and the dynamic, real-time data they encounter in practice. Offline models are typically built on datasets that capture past patterns of misinformation; however, as new events unfold and novel forms of fake news arise, the characteristics of these narratives may diverge substantially from those previously observed. This mismatch between the static training data and the constantly evolving nature of real-time news further aggravates the models' inability to adapt, leading to a decline in their detection accuracy when confronted with fresh and previously unseen misinformation.


Motivated by the identified gap in offline models' performance on real-time news, we explored the potential of online Large Language Models (LLMs) as a viable solution. Real-time fake news detection presents a new paradigm in the fight against misinformation, requiring systems that can continuously learn and adapt to emerging data streams. Unlike traditional models, LLMs are designed to process vast amounts of information in real-time and dynamically access credible online resources, enabling them to identify emerging patterns of misinformation with greater precision. Leveraging the advanced capabilities of LLMs, we aim to address the limitations of offline models, thereby enhancing the accuracy and responsiveness of fake news detection systems in today's fast-paced information environment.

In recent years, Large Language Models (LLMs) such as ChatGPT\footnote{ChatGPT: \url{https://openai.com/chatgpt}}, Claude\footnote{Claude: \url{https://claude.ai}}, Llama\footnote{Llama: \url{https://llama.meta.com}}, and Gemini\footnote{Gemini: \url{https://gemini.google.com}} have demonstrated remarkable performance across a wide range of natural language processing tasks, showcasing their potential to tackle complex challenges. Although the application of LLMs in real-time fake news detection has not been extensively explored, their ability to process and learn from real-time data suggests they may offer significant improvements in this domain. These models, trained on vast and diverse datasets, are capable of generating and understanding human-like text, and the ability to access online information, making them particularly well-suited for adapting to the constantly evolving landscape of online discourse. This adaptability positions LLMs as promising candidates for identifying new and emerging forms of fake news. Furthermore, their public accessibility, without requiring specialized technical expertise, enhances their potential for widespread use in fake news detection.

While LLMs demonstrate substantial potential, there is a pressing need to systematically evaluate their effectiveness in the context of real-time fake news detection, particularly in comparison to traditional models. Existing studies on LLMs for fake news detection \cite{hu2024bad, li2024large, boissonneault2024fake} primarily focus on fine-tuning these models using static, historical datasets before deploying them for fake news detection. Although this approach can enhance performance for specific tasks, it does not fully leverage LLMs' inherent ability to dynamically process real-time information. Moreover, these studies typically use models like GPT-3.5 Turbo, which do not have online search capabilities, restricting their ability to incorporate the most up-to-date information when evaluating news. This reliance on static data poses a challenge in the real-time context, where misinformation can spread rapidly, and models need to adapt to new narratives as they emerge.

In contrast, our work employs LLMs in their zero-shot capacity, without fine-tuning on task-specific datasets, allowing them to directly process and evaluate real-time news. This approach not only tests the models' adaptability to emerging narratives but also capitalizes on their ability to access up-to-date information, enabling more context-aware detection of fake news. By leveraging models with real-time web access, such as the latest versions of ChatGPT, Claude, and Gemini, we aim to overcome the limitations of static fine-tuned models. To our knowledge, this research is the first to rigorously test both LLMs and existing fake news detection models using real-time news datasets, offering a comprehensive evaluation that highlights the strengths and limitations of these approaches in handling real-time information.


Our study makes several key contributions:

\begin{itemize}
  \item \textbf{Identification of a research Gap}: We identify a significant shortfall in the ability of existing models to effectively address real-time fake news, highlighting the inadequacy of current detection systems.
  \item \textbf{Comprehensive evaluation on real-time news}: We provide a thorough comparison between LLMs and traditional offline models, elucidating their respective strengths and limitations in the context of real-time misinformation detection.
  \item \textbf{Insights for fake news detection}: Our findings offer valuable insights that can guide the development of more robust and adaptable fake news detection solutions, capable of handling the evolving nature of misinformation.
\end{itemize}

\section{Related Work}
The automatic detection of fake news on social media has become a significant research area due to the proliferation of misinformation and its impact on public opinion. Various approaches have been proposed over the years, which can be broadly categorized into machine learning methods, deep learning techniques, and multimodal approaches.

\subsection{Machine Learning Methods}
Traditional machine learning techniques have been widely used in the initial stages of fake news detection. These methods rely on handcrafted features extracted from the textual content of news articles. Logistic regression, support vector machines (SVM), and random forests are commonly employed classifiers in this category. For instance, Castillo et al. \cite{castillo2011information} developed a decision-tree-based model using features from Twitter events. Similarly, Rubin et al. \cite{rubin2015towards} utilized SVM combined with rhetorical structure theory and vector space modeling to classify news. Other researchers, such as Horne and Adali \cite{horne2017just}, have investigated various linguistic features, including stylistic features and complexity measures, to differentiate between fake and real news. They used SVM and achieved notable performance improvements by incorporating these features. Zhou et al. \cite{zhou2020fake} employed feature engineering techniques to extract stylistic and content-based features for fake news detection, demonstrating the effectiveness of logistic regression and random forests in this task. These models often utilize linguistic cues such as word n-grams, part-of-speech tags, readability scores, syntax patterns, and semantic inconsistencies to distinguish fake news from real news. The emotional tone of a piece of news can also serve as an indicator of its veracity, with fake news often exhibiting exaggerated sentiment compared to factual news. Additionally, assessing the credibility of the news source, including its history of publishing fake news and its overall reputation, can significantly enhance the accuracy of fake news detection models.

\subsection{Deep Learning Techniques}
Recent advancements have seen the application of deep learning models, which can automatically learn complex features from large datasets, significantly improving detection accuracy. Convolutional Neural Networks (CNNs), Recurrent Neural Networks (RNNs), and transformer-based models like BERT (Bidirectional Encoder Representations from Transformers) have been particularly effective. Wang \cite{wang2017liar} introduced the LIAR dataset and used various models including CNNs and Bi-LSTMs to analyze linguistic patterns in the data, with CNNs outperforming others. Kaliyar et al. \cite{kaliyar2021fakebert} proposed a transformer-based model named FakeBERT, which fine-tunes the BERT model for fake news detection, achieving state-of-the-art results on multiple benchmarks. Ni et al. \cite{ni2021mvan} created a model, called MVAN (Multi-View Attention Networks), based on deep learning to spot fake news on an early basis. They merged the text semantic attention and the propagation structure attention in the model in order to simultaneously gather important hidden cues from the originating tweet’s dissemination structure. These models leverage the powerful language representation capabilities of deep learning techniques to capture local textual features, long-term dependencies, and semantic relationships in the data, significantly improving the performance of fake news detection systems.

Hybrid models that combine different deep learning techniques have also been proposed. Ruchansky et al. \cite{ruchansky2017csi} developed the CSI (Capture, Score, and Integrate) model, which incorporates text, reaction, and source characteristics to detect fake news. Ajao et al. \cite{ajao2018fake} created a hybrid model based on LSTM and CNN for Twitter data. Nasir et al. \cite{nasir2021fake} developed a hybrid model based on deep learning for detecting fake news that uses recurrent and convolutional neural networks. These hybrid models combine the strengths of CNNs in extracting spatial features and LSTMs in capturing temporal sequences, providing a comprehensive approach to analyzing the complex nature of fake news.

\subsection{Multimodal Approaches}
Given the rise of multimedia content on social media, multimodal approaches that combine text and images have gained attention. Models integrating textual and visual information can provide a more comprehensive analysis of news content. Wang et al. \cite{wang2018eann} proposed EANN (Event Adversarial Neural Networks), which can learn event-invariant features from multimodal data to detect fake news, using adversarial learning to ensure that the learned features are robust across different events. Khattar et al. \cite{khattar2019mvae} introduced the MVAE (Multimodal Variational Autoencoder) for fake news detection, which uses variational autoencoders to capture the joint distribution of text and images, improving robustness and accuracy in fake news detection. Giachanou et al. \cite{giachanou2020multimodal} created a multimodal multi-image system that integrates text, visual, and semantic components to detect fake news. Their approach utilizes BERT for textual representation, VGG-16 for visual representation, and cosine similarity between text and image tags for semantic representation. Segura-Bedmar and Alonso-Bartolome \cite{segura2022multimodal} developed a multimodal fake news detection method that combines text and image data using a Convolutional Neural Network (CNN) architecture. These models leverage the complementary strengths of different modalities to provide a more holistic understanding of the news content, improving the accuracy and reliability of fake news detection systems.

\subsection{LLM-Based Fake News Detection}

The integration of large language models has significantly advanced the sophistication of fake news detection methods. Traditional approaches often rely on auxiliary data in addition to the article’s text. For example, detection systems like Grover \cite{zellers2019defending} use metadata such as author details, publishers, and publication dates to assess the authenticity of articles. Similarly, DeClare verifies the credibility of statements by comparing them against information gathered from web-searched articles \cite{popat2018declare}. And platforms like Defend analyze social media interactions with news articles to assist in fake news detection \cite{shu2019defend}. Additionally, methods such as those demonstrated by Zhang et al. in 2021 focus on extracting emotional and semantic features from texts to distinguish between real and fake news \cite{zhang2024mining}. These approaches, while effective, can be limited by the need for auxiliary data, which is not always readily available.

Recent studies have shifted towards using LLMs like GPT-3.5 for detecting fake news without the need for extensive auxiliary data. These models focus on analyzing the content of news articles themselves, utilizing their vast pre-trained knowledge to identify falsehoods more efficiently. In particular, fine-tuned versions of these models have shown strong performance on benchmark datasets. Models like FactAgent \cite{li2024large} propose advanced fact-checking workflows that enable the model to systematically decompose complex news claims into smaller tasks before verifying the veracity of each sub-claim. However, despite its innovative approach, the reliance on static datasets limits its adaptability in detecting misinformation in rapidly evolving real-time contexts. Similarly, HiSS (Hierarchical Step-by-Step Prompting) \cite{zhang-gao-2023-towards} employs a prompt-engineering strategy, where LLMs are guided to break down a claim into smaller sub-claims for verification using external evidence. 

Recent work by Hu et al. \cite{hu2024bad} explores the role of LLMs in fake news detection, finding that while LLMs like GPT-3.5 provide useful multi-perspective rationales, they often underperform in directly detecting fake news compared to fine-tuned small language models (SLMs). To address this, the authors propose an Adaptive Rationale Guidance (ARG) network, which leverages the strengths of both LLMs and SLMs, achieving improved performance. However, their approach remains focused on static datasets and does not fully explore the potential of LLMs in real-time news detection, a gap our work seeks to address. While these techniques improve the LLM's performance by structuring the fact-checking process, they do not address the challenges posed by real-time news, as they remain limited by the static data used during training.

A key distinction between earlier models like GPT-3.5 and the most current LLMs lies in their online capabilities. GPT-3.5 operates as an offline model, relying solely on static, pre-trained knowledge with no ability to access or retrieve live, real-time information from the web. In contrast, modern LLMs such as ChatGPT-4, Gemini, Claude, and Llama are designed with online access features, allowing them to dynamically update their knowledge and verify information in real-time. Given the increasing speed and sophistication of fake news dissemination, it is a natural progression to shift from offline models to these online-enabled models. This transition allows for a more adaptive and responsive approach to combating misinformation, making real-time fake news detection more efficient and scalable.

Despite these advancements, the potential of online LLMs for real-time fake news detection remains underexplored. Most existing models are evaluated on static, historical datasets and do not fully exploit the adaptability and real-time processing capabilities of LLMs. In contrast, our work takes a step further by evaluating LLMs in a real-time fake news detection context, without relying on extensive pre-training or task-specific fine-tuning.

Furthermore, the public accessibility of original LLMs, without requiring specialized technical expertise, enhances their potential for widespread use in fake news detection. Unlike many fine-tuned or task-specific models, which often require domain-specific data and technical resources to build, general-purpose LLMs are easily accessible to the general public. These models offer a flexible, scalable solution to identifying misinformation without the need for specialized knowledge in AI model fine-tuning. This underscores the importance of evaluating LLMs in the context of real-time fake news detection, as their potential reach far surpasses that of fine-tuned models.


\subsection{Challenges and Research Gaps}

Developing effective fake news detection systems involves several challenges. The evolving nature of fake news and the rapid dissemination of information on social media platforms require models that can adapt to new and emerging patterns of misinformation dynamically \cite{singh2024comprehensive}. Traditional models often struggle with real-time detection because they are trained on historical data and cannot easily incorporate new information.

A significant challenge is the need for models to process and analyze real-time data to identify fake news as it emerges. This requires the development of adaptive models that can learn from new data continuously and update their knowledge base. Additionally, there is a need to explore new datasets that better represent the current landscape of fake news, as existing datasets may become outdated quickly.

Due to these challenges and gaps, we aim to find robust and adaptable models that can handle the real-time nature of fake news. We want to explore the capabilities of LLMs like ChatGPT, Claude, Llama, and Gemini in detecting fake news. These models can process and learn from real-time data, potentially providing a more effective approach to identifying fake news as it appears.


\section{Limitations of Offline Models}
\subsection{Challenges in Real-Time News Detection}

Traditional fake news detection models are typically offline, meaning they are trained on historical datasets and do not adapt to new information in real-time. This static nature presents significant limitations, as fake news continuously evolves, with new stories and formats regularly emerging. Consequently, these offline models become less effective in identifying the latest misinformation, leading to decreased accuracy and relevance.

The effectiveness of a machine learning model is highly dependent on the quality of its training data. Generally, when training a model, it is assumed that the training dataset and the test/validation dataset follow the same or similar distribution. This assumption is crucial for the model to generalize well to new, unseen data. However, in the context of fake news detection, this assumption often does not hold. Real-time news can differ significantly from the historical data on which offline models are trained. If the validation dataset is not correlated to the training dataset, the trained model will struggle to perform effectively.

For instance, during the COVID-19 pandemic, an excessive amount of fake news stories emerged about the virus, vaccines, and treatments. Offline models trained on pre-pandemic data struggled to detect these new types of misinformation. An example includes false claims about the efficacy of certain treatments like hydroxychloroquine, which spread rapidly across social media platforms. Since offline models were not trained on such content, they often failed to identify these new falsehoods.

Another example is the spread of misinformation during significant political events, such as the 2020 US Presidential Election. Fake news articles and social media posts claiming election fraud or manipulating voting processes emerged quickly. Offline models, which did not have training data reflecting these specific events, showed limitations in accurately detecting and flagging such misinformation.

Additionally, offline models lack the flexibility to process and learn from new data dynamically. They require periodic retraining with updated datasets, which is both time-consuming and computationally expensive. This inflexibility further hinders their ability to stay current with the rapidly changing news landscape. For instance, the rapid evolution of fake news regarding new technologies, such as 5G networks causing health issues, posed a challenge to offline models that were not updated with the latest information.


\subsection{Experimental Validation of Offline Models on Real-Time News}

\subsubsection{Data Collection}

To evaluate the performance of LLMs and existing fake news detection models, we build a continuously updated dataset of real-time news. This dataset includes news articles posted in 2023 and 2024 from social media platforms like Twitter (X) and fact-checking websites such as PolitiFact. While each evaluation snapshot is static, the dataset itself remains dynamic, as we continuously incorporate the most recent news, ensuring it reflects the evolving nature of misinformation and real-time events.

The data collection process involves both automated and manual labeling to ensure accuracy and relevance. By regularly updating the dataset with fresh content from live online platforms, we simulate the conditions faced by real-time fake news detection systems. This dynamic nature allows us to assess the adaptability of models to new and emerging misinformation, differentiating our approach from prior studies that rely on outdated static datasets. Moving forward, we aim to further enhance this process by integrating live data feeds, enabling real-time testing environments.


\begin{table}[h!]
\centering
\renewcommand{\arraystretch}{1.5} 
\setlength{\tabcolsep}{12pt} 
\begin{tabular}{cc}
\hline
\textbf{News Type} & \textbf{Count} \\ \hline
Fake News         & 280           \\ 
Real News         & 55            \\ 
\textbf{Total}    & \textbf{335}  \\ \hline
\end{tabular}
\caption{Distribution of News in the Dataset}
\label{table:news_distribution}
\end{table}

\begin{table*}[]
\centering
\renewcommand{\arraystretch}{1.5} 
\setlength{\tabcolsep}{8pt} 
\begin{tabular}{|p{3cm}|p{13.5cm}|}
\hline
\textbf{Category}           & \textbf{Examples}                                                                                     \\ \hline
Politics     & The U.S. Supreme Court ruled in May that "Texas SB-4 law is constitutional."                     \\ \hline
Crime  &
  Anti-abortion activist sentenced to 57 months in prison for "handing roses and resources to women at an abortion facility." \\ \hline
Health &
  Infertility is treated differently than other issues and "often excluded from insurance coverage" \\ \hline
Entertainment                & Elon Musk fires entire cast of 'The View' after acquiring ABC.                                  \\ \hline
Economics                 & Florida has "the highest" homeowners insurance in the nation.                                   \\ \hline
Sports                & NFL referees "flex their authority, eject five players" for kneeling during the national anthem. \\ \hline
International Affairs &
  Israel "accidentally voted (for) Palestine" at a United Nations Security Council meeting. \\ \hline
Science                & NASA is "shooting three rockets at three moons" on the day of the solar eclipse.                \\ \hline
\end{tabular}
\caption{News Categories and Examples}
\end{table*}

\subsubsection{Models for Evaluation}

To evaluate the effectiveness of traditional fake news detection models in a real-time context, we selected three pre-trained models that do not require additional training. Although one of these models, ChatGPT-3.5 Turbo is an LLM, its offline nature—where it cannot access real-time data or updates—makes it suitable for comparison alongside other traditional offline models. The chosen models represent state-of-the-art approaches for detecting fake news across different domains and modalities:

\begin{itemize}
    \item \textbf{MDFEND (Multi-domain Fake News Detection)} \cite{nan2021mdfend}: Released in 2021, MDFEND  is designed to tackle the challenges of detecting fake news across multiple domains. MDFEND pre-trains a RoBERTa model on the Weibo21 dataset, which includes news from nine different domains, ensuring the model can adapt to various domains. 

    \item \textbf{Multimodal Fake News Detection} \cite{segura2022multimodal}: This model explores both unimodal and multimodal approaches for fake news detection using the Fakeddit dataset. The unimodal approaches include CNN, BiLSTM, and BERT, which focus solely on text data. 

    \item \textbf{Chatgpt 3.5Turbo} is a variant of OpenAI’s GPT-3.5 architecture, known for its efficiency in generating human-like text based on pre-trained data. Like other models in the GPT-3 series, it operates as an offline model, meaning that it does not have access to real-time data or the internet for live information retrieval. Instead, ChatGPT-3.5 Turbo generates responses based on the static dataset it was trained on, which contains information up until a certain cut-off date (typically 2021). Although ChatGPT-3.5 Turbo is a Large Language Model (LLM), its lack of real-time data access places it in a similar category to traditional offline models
    
\end{itemize}

While it is acknowledged that a broader comparison involving more benchmark methods would offer greater validation, several constraints influenced the selection process. Some existing methods require more detailed or specific information about the news content, such as user interactions or metadata, which were not available or applicable in our real-time dataset. Additionally, other state-of-the-art methods have not released their code or an accessible application, limiting their inclusion in this study. As a result, MDFEND, the Multimodal Fake News Detection model, and the ChatGPT-3.5Turbo were chosen for their accessibility and relevance, allowing for a focused yet insightful evaluation of offline models in the context of real-time news detection.

\subsubsection{Experimental Setup and Results}

To assess the performance of these offline models on real-time news, we conducted experiments using a dataset specifically designed to reflect the dynamic nature of contemporary misinformation. The models were evaluated across several key performance metrics: accuracy, precision, recall, and F1-score. As summarized in Table \ref{tab:metrics1}, while these models demonstrated strong performance on historical datasets, their effectiveness notably decreased when applied to real-time news. This decline in performance highlights the inherent limitations of offline models in adapting to the rapid evolution of fake news narratives.

\begin{table}[h]
\centering
\renewcommand{\arraystretch}{1.5} 
\setlength{\tabcolsep}{2pt} 
\begin{tabular}{ccccc}
\hline
\textbf{Model} & \textbf{Accuracy} & \textbf{Precision} & \textbf{Recall} & \textbf{F1 Score} \\ \hline
MDFEND & 0.837 & 0.492 & 0.641 & 0.557 \\ 
Multimodal  & 0.870 & 0.573 & 0.735 & 0.644 \\ 
ChatGPT-3.5Turbo  & 0.861 & 0.550 & 0.717 & 0.623 \\\hline
\end{tabular}
\caption{Performance metrics for selected pre-trained models evaluated on real-time news.}
\label{tab:metrics1}
\end{table}

The experimental results clearly underscore the challenges that offline models face in detecting fake news within a real-time context. Despite their strong performance on historical datasets, these models struggle to maintain accuracy and relevance when faced with new, emerging information. This significant performance drop reveals the static nature of offline models, which lack the ability to adapt to the constantly evolving landscape of fake news. These findings underscore the critical need for more adaptive and dynamic approaches to fake news detection, paving the way for the integration of online LLMs, which can process and learn from real-time data to offer a more robust and responsive solution.

\section{Online Large Language Models for Fake News Detection}

In contrast, LLMs like ChatGPT and Llama have shown great potential in various natural language processing tasks due to their extensive pre-training on vast datasets and fine-tuning capabilities. These models can understand and generate human-like text, making them highly adaptable to new information and contexts.

Moreover, online LLMs can access and process real-time data, providing a significant advantage over traditional models in the context of fake news detection. Their ability to continuously learn and update based on new information allows them to detect emerging fake news patterns more effectively. This real-time processing capability makes LLMs well-suited for the dynamic and fast-paced nature of fake news.

Existing comparative studies have demonstrated the superior performance of LLMs over traditional models in various NLP tasks, including sentiment analysis, language translation, and text generation. However, there is limited research specifically focused on evaluating LLMs for real-time fake news detection. Comparative studies that do exist highlight both the strengths and challenges of using LLMs for this purpose. While LLMs offer improved accuracy and adaptability, they also come with challenges such as ensuring the reliability of real-time data and managing the computational overhead associated with processing large volumes of data continuously.

The current state of research indicates a clear need for real-time fake news detection models. Traditional offline models, although useful, are limited by their reliance on outdated data and lack of adaptability to new information. Therefore, LLMs, with their ability to process and learn from real-time data, present a promising alternative. This study aims to bridge the gap in existing research by providing a comprehensive evaluation of LLMs for real-time fake news detection, highlighting their advantages and addressing potential challenges.

Additionally, our work focuses solely on detecting fake news based on plain news text, without requiring any additional resources such as user interactions or multimedia content. This approach ensures that the models are evaluated purely on their ability to process and understand textual information, providing a clear comparison between traditional models and large language models in a real-time news detection context.

\subsection{LLM Models for Evaluation}
Large Language Models:

\begin{enumerate}
    \item \textbf{ChatGPT (4o):} ChatGPT-4o is the latest version of the ChatGPT model developed by OpenAI, released in May 2024. This cutting-edge LLM is renowned for its conversational abilities and extensive pre-training on diverse datasets. It is designed to process real-time data efficiently, making it highly effective for detecting fake news. Its conversational nature allows it to understand and generate human-like text, enhancing its capability to identify misinformation.

    \item \textbf{Claude (3.5 Sonnet):} Claude-3.5 Sonnet , developed by Anthropic, was released in June 2024. Claude focuses on delivering accurate and context-aware responses and it has been pre-trained on a wide range of information, making it adept at identifying fake news in real-time scenarios. Claude's strength lies in its detailed and nuanced understanding of context, which is crucial for discerning misinformation.

    \item \textbf{Llama (3.1-405B):} Llama-3.1 \cite{dubey2024llama}, developed by Meta AI, was released in July 2024. Llama is known for its adaptability and efficiency in processing large volumes of text and it benefits from continuous learning and real-time data access, enhancing its fake news detection capabilities. Llama's design focuses on scalability and responsiveness, making it an excellent tool for handling the dynamic nature of real-time news.

    \item \textbf{Gemini (1.5 Pro):} Gemini-1.5 Pro is developed by Google DeepMind, released in June 2024. Gemini excels in natural language understanding and generation, and its continuous learning ability from new data sources and frequent updates makes it well-suited for real-time fake news detection tasks. Gemini's robust architecture ensures it stays current with evolving news patterns, providing accurate and timely responses.
\end{enumerate}

These LLM models, all of which are the latest versions available at the time of the experiment, will be evaluated on their ability to accurately and efficiently detect fake news in a real-time context. The comparison aims to highlight the strengths and weaknesses of traditional models versus LLMs. Using the latest versions ensures that we leverage the most advanced capabilities and improvements, providing a clearer picture of the current state-of-the-art in fake news detection. This evaluation will offer insights into the effectiveness and practicality of using LLMs for this critical task, especially in dynamically changing environments. Table \ref{table:llms_overview} displays an overview of the leading large language models (LLMs).

\begin{table*}[]
\centering
\renewcommand{\arraystretch}{1.5} 
\setlength{\tabcolsep}{8pt} 
\begin{tabular}{p{3cm}p{3cm}p{2cm}p{7.5cm}}
\hline
\textbf{Model} & \textbf{Developer} & \textbf{Release Date} & \textbf{Key Features} \\ \hline
\textbf{ChatGPT (4O)} & OpenAI & May 2024 & Conversational abilities, extensive pre-training on diverse datasets, efficient real-time data processing, effective for detecting fake news \\ 
\textbf{Claude (3.5 Sonnet)} & Anthropic & June 2024 & Accurate and context-aware responses, pre-trained on a wide range of information, detailed understanding of context for misinformation detection \\ 
\textbf{Llama (3.1)} & Meta AI & July 2024 & Adaptability and efficiency in processing large text volumes, continuous learning and real-time data access, scalable and responsive design for handling dynamic real-time news \\ 
\textbf{Gemini (1.5 Pro)} & Google DeepMind & June 2024 & Natural language understanding and generation, continuous learning from new data sources, frequent updates, robust architecture for real-time fake news detection \\ \hline
\end{tabular}
\caption{Overview of LLMs for Fake News Detection}
\label{table:llms_overview}
\end{table*}

\subsection{Evaluation Framework}

The ability of LLMs to detect real-time fake news was evaluated using the same real-time news dataset used to evaluate existing models. This section details the experimental setup, including the zero-shot approach employed for the evaluation of LLMs.

Given the exploratory nature of this research, we adopted a zero-shot evaluation approach for the LLMs, which allows these models to be tested on tasks without explicit prior training on similar datasets. The LLMs used in this study, including ChatGPT, Claude, Llama, and Gemini, were evaluated by posing them the task directly: ``I will give you some news; please determine whether it is real news or fake news.'' This approach mirrors real-world applications where LLMs are often employed without fine-tuning on specific datasets, relying instead on their broad, pre-trained, and fresh online knowledge.

During the evaluation, LLMs occasionally responded with phrases such as "highly likely to be fake" or ``highly likely to be true.'' For the purposes of this study, responses indicating a high likelihood of being fake were treated as ``fake,'' while those indicating a high likelihood of being true were treated as ``true.'' In instances where the LLM responded with uncertainty, such as ``I don't know,'' this was treated as a failed response. The rationale behind this approach is that an inability to definitively classify the news reflects a failure to detect the fake news, which is crucial in the context of real-time detection.

The performance of both the traditional models and LLMs was evaluated using several key metrics: accuracy, precision, recall, and F1-score. These metrics provide a comprehensive view of each model’s ability to correctly identify fake news while minimizing false positives and false negatives. The results from these evaluations are discussed in detail in the subsequent sections, highlighting the strengths and limitations of each approach within the context of real-time news detection.

\subsubsection{Experimental Results}

The experimental evaluation was conducted to assess the performance of four LLMs (i.e., ChatGPT, Claude, Llama, and Gemini) on real-time fake news detection tasks. These models were evaluated using key performance metrics, including accuracy, precision, recall, and F1-score. The overall performance results are summarized in Table \ref{tab:performance_models}.

\begin{table}[h]
\centering
\renewcommand{\arraystretch}{1.5} 
\setlength{\tabcolsep}{5pt} 
\begin{tabular}{ccccc}
\hline
\textbf{Model} & \textbf{Accuracy} & \textbf{Precision} & \textbf{Recall} & \textbf{F1 Score} \\ \hline
ChatGPT & 0.946 & 0.897 & 0.909 & 0.903 \\ 
Claude  & 0.922 & 0.930 & 0.778 & 0.829 \\ 
Llama   & 0.949 & 0.888 & 0.947 & 0.914 \\ 
Gemini  & 0.875 & 0.590 & 0.787 & 0.675  \\ \hline
\end{tabular}
\caption{Performance metrics for tested LLM models.}
\label{tab:performance_models}
\end{table}

The results indicate that Llama achieved the highest overall accuracy at 0.949, followed closely by ChatGPT at 0.946, with both models demonstrating strong performance across all metrics. Claude, while achieving the highest precision at 0.930, displayed a trade-off with a lower recall of 0.778, suggesting that it may miss some instances of fake news. Gemini, although showing potential, exhibited lower performance metrics overall, with an F1-score of 0.675, highlighting the challenges it faces in this task.

Llama's performance was particularly notable in terms of recall, achieving a score of 0.947, indicating its robustness in detecting a wide range of fake news instances. On the other hand, ChatGPT offered a balanced performance with an F1-score of 0.903, making it a well-rounded option for real-time detection scenarios. Claude, despite its precision, had a lower F1-score of 0.829 due to its comparatively lower recall. These findings underscore the varying strengths and trade-offs among the models, which will be further examined in the discussion section.

In addition to overall performance, we also evaluated the models' effectiveness across different domains, including Politics, Crime, Health, Entertainment, Economics, Sports, International Affairs, and Science. The domain-specific performance, measured by F1-score, is presented in Table \ref{tab:performance_domain}.

\begin{table*}[]
\centering
\renewcommand{\arraystretch}{1.5} 
\setlength{\tabcolsep}{6pt} 
\begin{tabular}{ccccccccccc}
\cline{1-9}
Model   & Politics & Crime & Health & Entertainment & Economics & Sports & International Affairs & Science &  \\ \cline{1-9}
ChatGPT & 0.769         & 1.000       & 0.875        & 1.000        & 0.947        & 0.888        & 0.800         & 0.800          \\
Claude  & 0.774         & 0.857       & 0.933        & 0.952        & 0.750        &  0.857       & 0.800         & 0.800          \\
Llama   & 0.850         & 0.800       & 1.000        & 0.909        & 0.900        & 1.000        & 0.857         & 0.857          \\ 
Gemini  & -             & -           & 0.778        & 0.952        & 0.823        & 0.833        & 0.444         & 0.667          \\\cline{1-9}
\end{tabular}
\caption{Domain-specific fake news detection performance (F1-score).}
\label{tab:performance_domain}
\end{table*}

During the evaluation, Gemini exhibited challenges when tasked with detecting fake news in the Politics and Crime domains. Specifically, the model often responded with statements such as ``I can't help with responses on elections and political figures right now. While I would never deliberately share something that's inaccurate, I can make mistakes. So, while I work on improving, you can try Google Search.'' As a result, we did not list the performance of Gemini under these two domains. This limitation reflects the model’s current restrictions in handling certain sensitive topics, which impacts its overall effectiveness in those areas.

These detailed domain-specific results highlight the varying capabilities of LLMs in detecting fake news across different content areas. This analysis provides a nuanced understanding of where each model excels and where improvements may be needed, offering valuable insights for the development of more specialized and effective real-time detection systems.

\section{Discussion}

The results of our experimental evaluation offer insights into the limitations of traditional offline models and the advantages of LLMs in real-time fake news detection. This section explores these findings, emphasizing the areas where LLMs excel and their potential for advancing misinformation detection.

Our study revealed substantial limitations in the ability of traditional offline models, such as MDFEND and the Multimodal Fake News Detection model, to adapt to the rapidly evolving nature of real-time news. These models were originally designed to perform well on static datasets, where the characteristics of the data remain relatively consistent over time. However, when applied to real-time news, where new narratives and misinformation patterns can emerge unpredictably, these models struggle to maintain accuracy and relevance. This challenge is particularly pronounced due to the distributional shift between the static training data and the dynamic, real-time data that these models encounter in practice. As fake news evolves in both content and dissemination tactics, the static nature of offline models becomes a significant drawback. Lacking the ability to continuously learn from new data streams, these models are less effective at detecting emerging forms of misinformation that deviate from the patterns observed during training. This limitation was evident in the lower performance metrics observed during our evaluation, particularly when contrasted with the more adaptive capabilities of LLMs.

In contrast, LLMs demonstrated a robust ability to manage the complexities of real-time fake news detection. The zero-shot evaluation approach employed in this study allowed us to assess the inherent capabilities of LLMs without the need for task-specific fine-tuning. The results showed that LLMs like ChatGPT, Claude, Llama, and even Gemini, despite some limitations, can achieve high accuracy and F1-scores, effectively handling novel and evolving information.

One of the key advantages of LLMs is their ability to provide nuanced and context-aware responses, unlike traditional models which are typically restricted to binary outputs—classifying news as either true or false. LLMs, on the other hand, can offer more detailed assessments. For instance, in evaluating the claim that "The 9th Circuit Court of Appeals just ruled Covid vax mandates unconstitutional," an LLM like ChatGPT not only identified the claim as false or misleading but also provided a detailed analysis. The model highlighted specific reasons for its conclusion, such as the lack of recent rulings, the nuanced nature of past court decisions, and the misleading use of terms like "just ruled." Additionally, the LLM flagged red flags indicating the news item’s likely falsehood, including oversimplification of legal issues, lack of specific details, and the absence of corroboration from credible sources. This level of detail and context-aware assessment goes beyond what traditional models can offer, providing users with a richer understanding of why a piece of news might be considered false.

Moreover, LLMs can leverage their extensive pre-training on diverse datasets to understand and analyze the context of the news, considering factors such as the source, the language used, and the broader socio-political environment in which the news is situated. This enables LLMs to detect more subtle forms of misinformation that may not be immediately apparent to traditional models. For example, an LLM might recognize that a piece of news is highly likely to be fake based on its resemblance to previously debunked stories, even if the specific content is new.

Another significant advantage of LLMs is their ability to generate explanations for their classifications. In our experiments, LLMs occasionally provided reasoning behind their predictions, such as pointing out inconsistencies in the narrative or referencing known facts that contradict the news content. This explanatory capability is a powerful tool for users who need to understand not just whether news is fake, but why it might be considered so. The ability to generate such explanations not only aids in enhancing user trust but also contributes to the broader effort of improving the transparency and accountability of AI systems in sensitive tasks like misinformation detection.

However, it is important to note the limitations faced by certain LLMs, such as Gemini, which exhibited challenges when dealing with politically sensitive topics. The model's tendency to respond with disclaimers, such as its inability to provide assistance with political content, highlights the need for further refinement in handling a broader range of topics, especially those that are critical in the context of misinformation.

In summary, while traditional offline models face significant challenges in adapting to the real-time detection of fake news, LLMs present a promising alternative with their adaptability, nuanced understanding, and ability to provide context-aware responses. These findings suggest that LLMs could play a pivotal role in advancing the field of misinformation detection, particularly in real-time applications where the ability to quickly and accurately identify false information is crucial.

\section{Conclusion}

The findings from this study suggest that LLMs hold significant promise for enhancing real-time fake news detection. Their adaptability, contextual awareness, and ability to provide detailed and nuanced assessments make them a valuable tool in the fight against misinformation. However, it is also important to recognize that while LLMs performed well in this study, their application in real-world scenarios will require careful consideration of factors such as computational resources, deployment costs, and the need for continuous updates to maintain their effectiveness.

Future research should explore the integration of LLMs with traditional models, leveraging the strengths of both approaches to create hybrid systems that can provide both high accuracy and contextual explanations. Additionally, there is a need for ongoing evaluation of LLMs in diverse and rapidly changing information environments to ensure that they continue to perform effectively as new forms of misinformation emerge.

In conclusion, while traditional offline models face significant challenges in adapting to real-time fake news detection, LLMs offer a promising alternative that combines high performance with the ability to provide richer and more informative outputs. As the landscape of misinformation continues to evolve, the deployment of LLMs in this context could play a crucial role in mitigating the spread of fake news and enhancing the quality of information available to the public.

\section*{Acknowledgment}

\bibliographystyle{IEEEtran}
\bibliography{cite.bib}

\vspace{12pt}

\end{document}